\documentclass[proceedings, preprint]{rmaa}

\SetYear{2022}
\SetConfTitle{Prospects for low-frequency radio astronomy in S. A.}

\title{Science research from the Instituto Argentino de Radioastronomía} 

\author{ P. Benaglia\altaffilmark{1} }

\altaffiltext{1}{Instituto Argentino de Radioastronomía (CONICET -- CIC -- UNLP), C.C.5, (1984) Villa Elisa, Buenos Aires, Argentina (paula@iar-conicet.gov.ar).}

\shortauthor{Benaglia}
\shorttitle{Science research from IAR}

\listofauthors{P. Benaglia}
\indexauthor{Benaglia, P.}

\abstract{In this talk, I will present some figures and milestones of the written production of the Instituto Argentino de Radioastronomía (IAR), as well as a personal review of the scientific achievements carried out in recent years by the researchers working at the IAR. I will also briefly describe the scientific objectives of the IAR's flagship project, the Multipurpose Interferometric Array (MIA), in the context of the instrumental projects that have lately been or are being installed on Argentine soil.}

\resumen{En esta charla, presentaré algunos números e hitos de producción escrita desde el Instituto Argentino de Radioastronomía (IAR), y una revisión personal de logros científicos llevados a cabo en los últimos años por investigadores que trabajan en el IAR. También, y en el marco de los proyectos instrumentales recientemente instalados -- o en vías de serlo -- en suelo argentino, describiré sucintamente los objetivos científicos del proyecto insignia del IAR, el Multipurpose Interferometric Array o MIA (por sus siglas en inglés).}

\addkeyword{Publications, bibliography}
\addkeyword{history and philosophy of astronomy}
\addkeyword{miscellaneous}

\begin{document}
\maketitle

\section{Introduction}
\label{sec:intro}

The Instituto Argentino de Radioastronomía (IAR) was founded in October 1962, originally under the name Instituto Nacional de Radioastronomia; for details see \citet{Romero2023}, this volume. Less than three years later, in April 1965, its 30-m single dish radiometer detected for the first time the neutral hydrogen (HI) line at 1.4~GHz. Observations of this transition were the strongest driver for building an observatory at our latitudes, since HI was discovered as a tracer of Galactic structure tracer, and an important part of the southern sky was inaccessible from most of the radio telescope sites operating at the time (in the northern hemisphere). 

The first paper published with IAR affiliation was \citet{Varsa1966}. And the first publication using data obtained from observations with the IAR's first radio telescope was that of \citet{Mesza1968}. 

\section{Some numbers and milestones}
\label{sec:milestones}

\begin{figure*}[!t]
   \includegraphics[width=16.5cm]{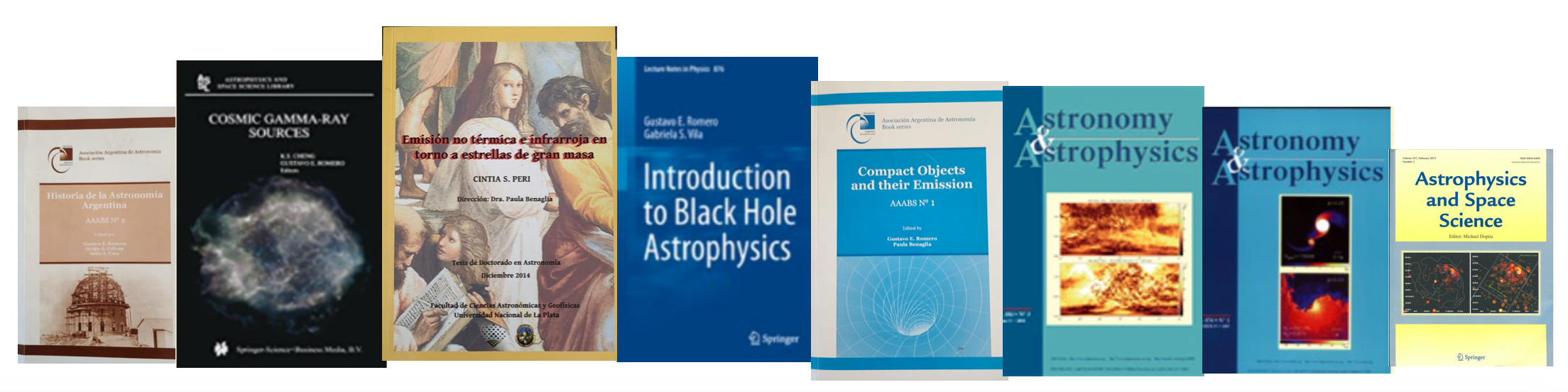}
  \caption{Some publications including books, dissertations, proceedings, cover page articles in peer-reviewed journals, etc.}
  \label{figure1}
\end{figure*}

In the 60~yr of its existence, about 1500 works have been published by authors affiliated to the IAR. The number includes articles in periodic journals, proceedings of professional meetings, theses at all levels, technical reports, books (Fig.~\ref{figure1}). 

The production of theses includes doctoral theses and `licenciatura' thesis (equivalent to a Bachelor's and Master's degree in Argentine National Universities), among other formats. 
The works correspond mainly to presentations given at the National University of La Plata and the University of Buenos Aires. In terms of awards, the doctoral theses developed at the IAR received three times the prize for the best theses of the biennium, awarded by the Asociación Argentina de Astronomía (www.astronomiaargentina.org), and one of the equivalent (Giambiagi prize) of the Argentine Physics Association. In 2020, a doctoral thesis conducted both at the IAR and the Karlsruhe Institute of Technology became the first one in the institution to be completed as a double doctorate (University of Karlsruhe and National University of San Martín), and the volume was selected by Springer as one of the world's best theses of the year and published in book form in Springer's ``Great Theses'' series.

Several articles first authored by IAR researchers have appeared on the front pages of major journals, others have received honors from the Gravity Research Foundation, and once even the Top Scientific Contribution Award, from the American Institute of Physics as one of the most cited papers of the year in The Astrophysical Journal.

The IAR's first 30-m dish radiometer - christened Carlos M. Varsavsky, as its first Director - was the second instrument of its kind to operate systematically in the southern hemisphere, along with the Parkes radio telescope in Australia. This led to many discoveries in the southern sky. Together with the second radio telescope to be built, now called Esteban Bajaja and completed around 1980, both continuum and line observations could be carried out, mainly at 1.4~GHz (atomic hydrogen), and 1.6~GHz (CO molecular transition).

Until the nineties, a great deal of time was devoted to mapping the distribution and velocity of the HI. \citet{Colomb1980} (CPH) presented images and profiles covering positions with declination ${\delta \leq 30^{\arcdeg}}$, each $1^{\arcdeg}$.
The CPH survey complemented that of \citet{Heiles1974}, which was carried out using the Hat Creek radio telescope, with 28-m dish. The sensitivity of the data was between 0.1 and 1~K.

After a change of receivers, among other things, the southern sky was again surveyed at the 1.4~GHz center frequency, with higher sensitivity: the system temperature of $\sim$25~K allowed an rms noise below 0.01~K. \citet{bajaja2005}, and \citet{kalberla2005} present the HI distribution and kinematic information, complementing the work of \citet{harbur1997}. The continuum emission with full polarization information has also been surveyed, but with the latest radiometer, achieving a sensitivity of 15~mK in the Stokes parameters Q and U \citep{testori2008}, for the coordinates not covered by \citet{reichreich1986}.

Results on certain, in some respects special, sources were achieved. For example, with the radio telescopes of the IAR the passage of the tail of the comet Halley was observed \citep{bajaja1987} 4~h a day for 3 months in 1986 to measure OH lines in absorption. The data allowed to derive the abundance of the molecule and the OH production rate. HI observations led to the discovery of the supernova remnant Vela Jr \citep{combi1999}. The extreme variability of Active Galactic Nuclei of blazar type was reported for the first time, after being measured with one of the IAR dishes \citep{romero1994}. In 2010, the first stellar bow shock with evidence of non-thermal emission was discovered using the Very Large Array, with a follow-up study of the polarization of the emission \citep{paula2021}. The Long Baseline Australian Array was used to map the fifth - at that time - colliding wind region of a massive binary stellar system (HD~93120AaAb) and to estimate important stellar/wind parameters \citep{paula2015}. \citet{manuel2014} published results of long-term mosaic  observations of the Serpens South molecular cloud with the Combined Array for Research in Millimeter-wave Astronomy, revealing new features in the gas dynamics.

\section{Main lines of research}
\label{sec:researchlines}

The research done at the IAR, let's say in the last decade or so, covers several fields, in astrophysics, physics, computer science, mathematics, etc.

There is a lot of activity in the area of extragalactic sources, such as AGN and their (super)massive black holes, extended to gravitational waves, cosmic rays, sources studied by means of electromagnetic cascades, neutrino astrophysics, field theory and relativity. 

Compact objects such as neutron stars or black holes, X-ray binary studies, have a common ground with those of HE (gamma-ray) sources or other candidates to produce HE radiation (i.e. colliding wind binaries, stellar bow shocks).

Stellar objects are widely studied, including their parent molecular clouds, star formation scenarios both of massive and low-mass types, interaction with the interstellar medium (ISM) and the ISM itself, the supernova stage and remnants, and planetary science.

The studies include the construction of mathematical models, add numerical simulations, and signal processing algorithms.

In the following, examples of research at the IAR are presented as divided in three groups: those carried out with the IAR radio telescopes, those where theoretical developments dominate, and those where observations with instruments around the world play a key role. 

\section{Research with IAR radiometers}

After about fifteen years out of use, the radio telescopes at the IAR have been refurbished \citep{gancio2020}, especially challenged by a major project to study transients, meaning by that sources that experiment changes in radiation with duration of the order of a second of time or much less. The observations resumed in 2018, and the data collection as a regular operation started in 2019. The mentioned project is organized in the framework of a scientific and technological cooperation between IAR and the Rochester Institute of Technology (USA). The group involved at IAR is called Pulsar Monitoring in Argentina (PuMA) \citep[see][this volume]{lousto2023}. 

The updated radio telescopes have digital back-ends (CASPER ROACH cards with 4x400-MHz of bandwidth). These are capable of recording two polarizations at 1420 MHz. They can be remotely controlled, and have access to an atomic clock, GPS and GNSS for timing purposes. Sources with declinations less than $-7{\arcdeg}$ can be tracked for 4 h per day.
 
The new architecture is optimal for studies involving pulsar timing arrays, targeted searches for continuous gravitational wave sources, monitoring of magnetars and glitching pulsars, and short time scale interstellar scintillation. The timing precision is better than 1$\mu$s \citep[e.g.][this volume]{zubieta2023}.

Another advantage is the geographical location of the IAR: sources that are invisible from Australia at 12~h daily interval and from South Africa at 5~h daily interval can be accessed with the IAR radio telescopes. In this way, alerts of detections between these three places and their instruments allow to follow a transient phenomenon along its entire occurrence.
 
\section{Theory, simulations, models}

\begin{figure}[!t]
  \includegraphics[width=7cm]{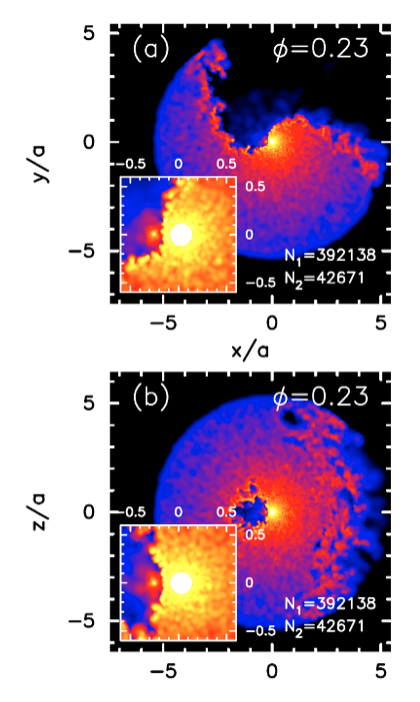}
  \caption{Wind collision interface geometry for the pulsar case in the orbital plane (top) and in a perpendicular plane (bottom), from \citet{romero2007}.}
  \label{figure2}
\end{figure}

With the creation of the Group of Relativistic Astrophysics and Radio Astronomy (GARRA, by its Spanish acronym), the scientific work at IAR was given a strong capacity for the development of theoretical studies. Articles such as \citet{romero2007}, chosen as the cover of the journal Astronomy and Astrophysics,   
are good first examples of such research. The authors focused on a challenging gamma-ray binary (then usually referred to as a microquasar), LS~I~$+$61~303, a Be-star with a compact  companion, to elucidate the nature of the compact object, between pulsar-neutron star or black hole. The light curve, the spectrum of the observed TeV gamma-ray emission, and the required energy budget were analyzed. In particular, they modeled the interaction between the components of the system for both hypotheses (Fig.~\ref{figure2}). For this, they obtained time on the powerful supercomputer HITACHI SR11000 at Hokkaido University. The results were consistent with the second case. 

Detailed models of the spectral energy distribution of massive binary systems (colliding-wind binaries) and the interaction with the interstellar medium were also developed; see \citet{maria2012}, \citet{santi2018}, \citet{santi2022}.

In \citet{vieyro2019} the authors studied the case of a core-collapse supernova inside the relativistic jet of an active galactic nucleus. After the analyzing the dynamical evolution of the supernova ejecta impacted by the jet and the observed gamma-ray light curve, they computed the spectral energy distribution for two different galaxy hosts, a radio galaxy and a blazar. They concluded that the first scenario appears to be much more common and easier to detect than the other.

Collaborations on planet formation are presented in \citet{irina2019}. The article deals with modelling the fragmentation of planetesimal in the formation of giant planets, taking into account the relative velocities and compositions of the planetesimals and the accretion produced by their fragmentation.

An example of progress in cosmology is presented in \citet{dperez2022} and its references, related to a universe with contractions and bounces. The survival of certain structures along them, in this case black holes, is studied using a generalized McVittie's metric.   

\citet{garciafede} studied physical and geometrical properties of the corona of the microquasar GRS~1915$+$105. They applied a variable comptonization model vKompth, -- developed by  IAR's Ph.D. student C. Bellavita \citep{candela} --, supported by archival X-ray observations, and found consistent trends in the evolution of the corona size, temperature, and feedback \citep[see also][]{mmendez2022}.

\begin{figure}[!t]
  \includegraphics[width=7.8cm]{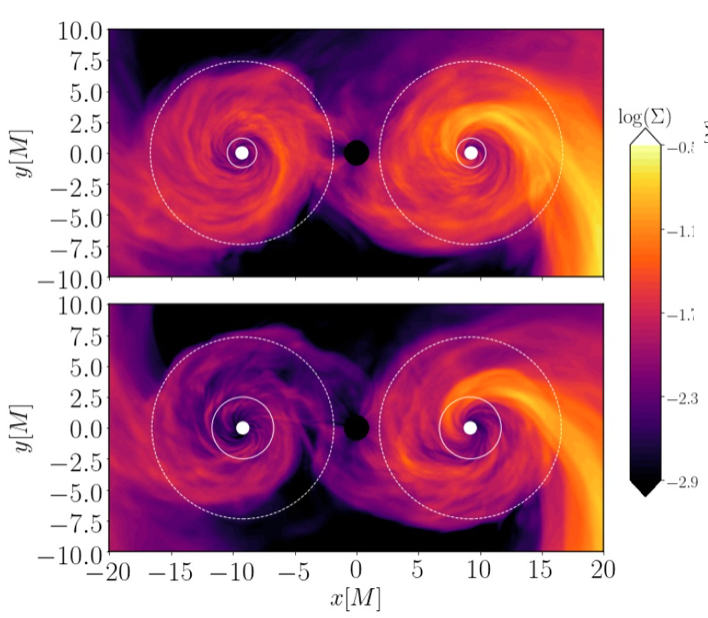}
  \caption{Surface density snapshot of a binary black hole system for two epochs \citep{lcombi2022}.}
  \label{figure3}
\end{figure}

As can be seen, Ph.D. theses are of particular interest at the IAR. For example, L. Combi studied binary black holes - in principle of equal mass, spinning, and approaching merger. Three-dimensional general relativistic magnetohydrodynamical simulations were performed on a system geometry that included a circumbinary disk and mini-disks around each black hole. These allowed to study the gas dynamics and system evolution, the morphology and variability of the electromagnetic flux densities, and to analyze the accretion (see Fig.~\ref{figure3}). The results on realistic synthetic light curves and spectra are very valuable for future observations with instruments like the Laser Interferometer Space Antenna (LISA) \citep{lcombi2022}.

\begin{figure}[!t]
  \includegraphics[width=7.8cm]{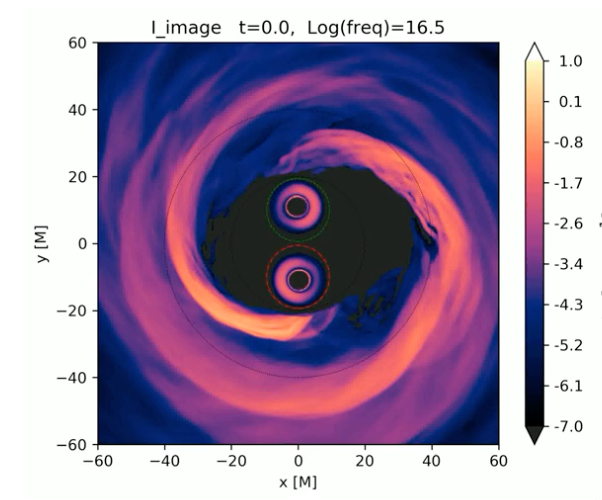}
  \caption{Stokes I instantaneous radiation distribution of a binary supermassive black hole approaching periastron \citep{edugut2022}.}
  \label{figure4}
\end{figure}

In the same subject and with a similar treatment, E.~M. Gutiérrez Ph.D. project consisted more in the study of the radiation from supermassive binary black holes \citep{edugut2022}. Simulations including blackbody radiation from an optically thick accretion disk, and hard X-rays from optically thin corona allowed to obtain spectra, images and light curves (Fig.~\ref{figure4}).

The Ph.D. Thesis by F. Fogantini was  focused on the phenomenology of accreting high-mass X-ray binaries (HMXBs). In this Thesis, it was demonstrated how geometrical or line-of-sight effects have a strong impact on the observed spectral and variability properties of archetypical HMXBs sources like SS 433 \citep{fogantini}.

\section{Observing with interferometric arrays}

The regular acquisition of observing time with instruments in foreign soil started around 2000, and is growing steadily mainly pursued by members of the fringe research group (Formation in Radio Interferometry - arGEntina). 

\begin{figure}[!t]
  \includegraphics[width=7.8cm]{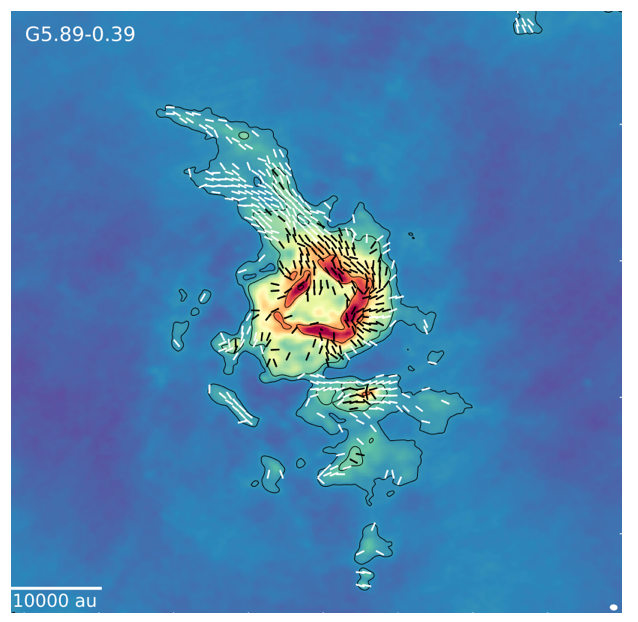}
  \caption{Electrical vector position angle rotated 90$\arcdeg$,  
  superpimposed on Stokes I continuum emission towards G5.89--0.39 \citep[full details in][]{manuel2021}.}
  \label{figure5}
\end{figure}

In recent years, the field of high-mass star formation has been shaken by the discovery of explosive outflows, in addition to the well-known bipolar outflows; \citep[see][and references therein]{zapata2009}. Only a few such regions have been described. One is G5.98--0.39, an ultra-compact HII region with formation of massive stars. \citet{zapata2020} and \citet{manuel2021} spotted it with the Atacama Large Millimeter/submillimeter Array (ALMA), and found dozens of CO filaments, and expanding warmer SiO gas at the origin. The energy released was inferred to be $\sim10^{46}-10^{49}$~erg. There is a north-south filamentary structure, a compact HII region, and a possibly expanding dusty belt, which harbours an O5V star. Polarized emission could be measured in the filaments, $\sim$4.4\%, coming from magnetically aligned dust grains. As Fig.~\ref{figure5} shows, the magnetic field lines in the central belt of dust are radially aligned.

\begin{figure}[!t]
  \includegraphics[width=7.8cm]{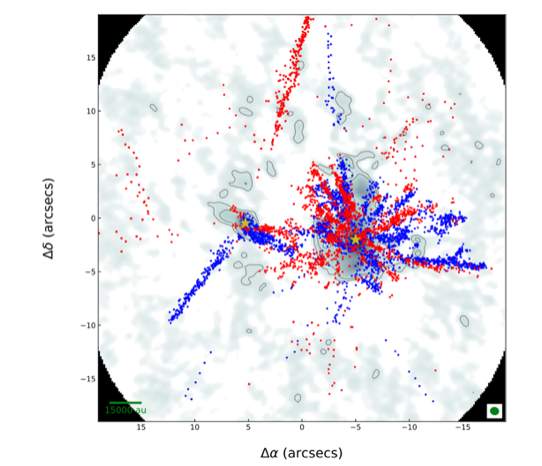}
  \caption{Red-shifted and blue-shifted condensations in IRAS~16076--513 \citep{estrella2022}.}
  \label{figure6}
\end{figure}

In \citet{estrella2022}, part of first author's Ph.D. thesis is presented, on the object IRAS~16076--5134, a high-mass star-forming region studied with ALMA band-7 CO archive data. Fourteen cores were detected. The imaged morphology and kinematics suggest a dispersal, explosive outflow, with filament-like CO ejections from a central position (see Fig.~\ref{figure6}), quasi isotropic; several filaments show a linear velocity gradient.

\begin{figure}[!t]
  \includegraphics[width=7.8cm]{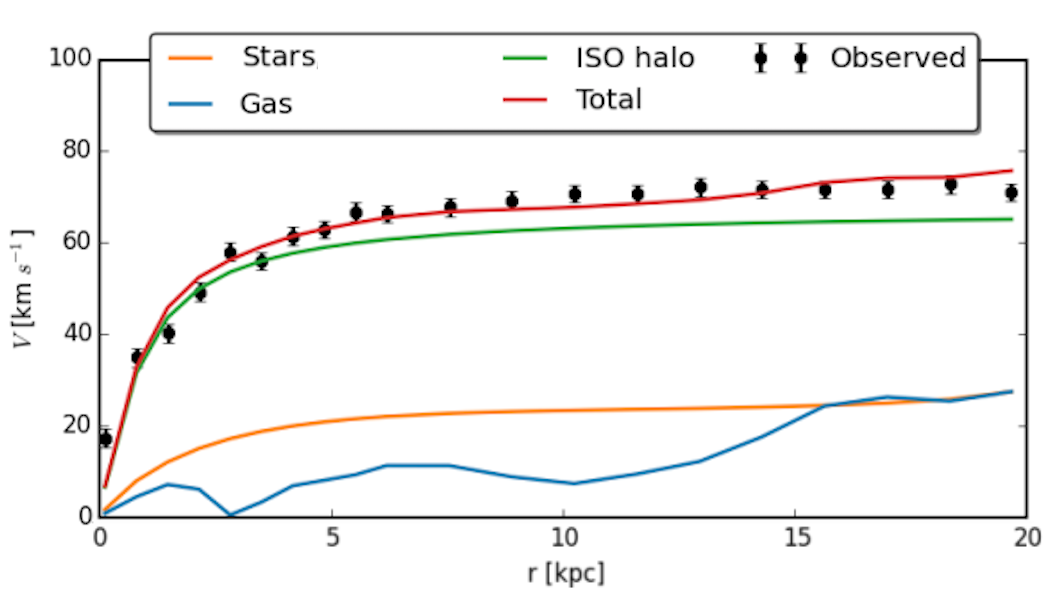}
  \caption{HI rotation curve model of the Fourcade-Figueroa galaxy. The orange line shows the rotation curve due to the stellar disk, the blue line shows the contribution due to the gas disk, the green line shows that due to the dark matter halo and the red colour shows the best-fitting model rotation curve. See details in \citet{jota2021}.}
  \label{figure7}
\end{figure}

Another example of recent Ph.D. project was the one focused on galaxy groups and certain types of galaxies (dwarf-, low surface brightness-, super thin-, local galaxies). The groups are a very common environment, easy to study because of the low relative velocities involved. The sources are studied by means of HI-line data. Questions such as which is the role of the environment in galaxy evolution, or how galaxy mergers, ram-pressure stripping, gravitational interactions and intragroup medium affect star formation and morphology are investigated. One of the contributed papers \citep{jota2021}, aimed at the superthin galaxy Fourcade-Figueroa, consisted in modeling the HI distribution, deriving the rotation curve, to finally obtain, also through modeling, characteristics of the dark matter halo (see Fig.~\ref{figure7}).

\begin{figure}[!t]
  \includegraphics[width=7.8cm]{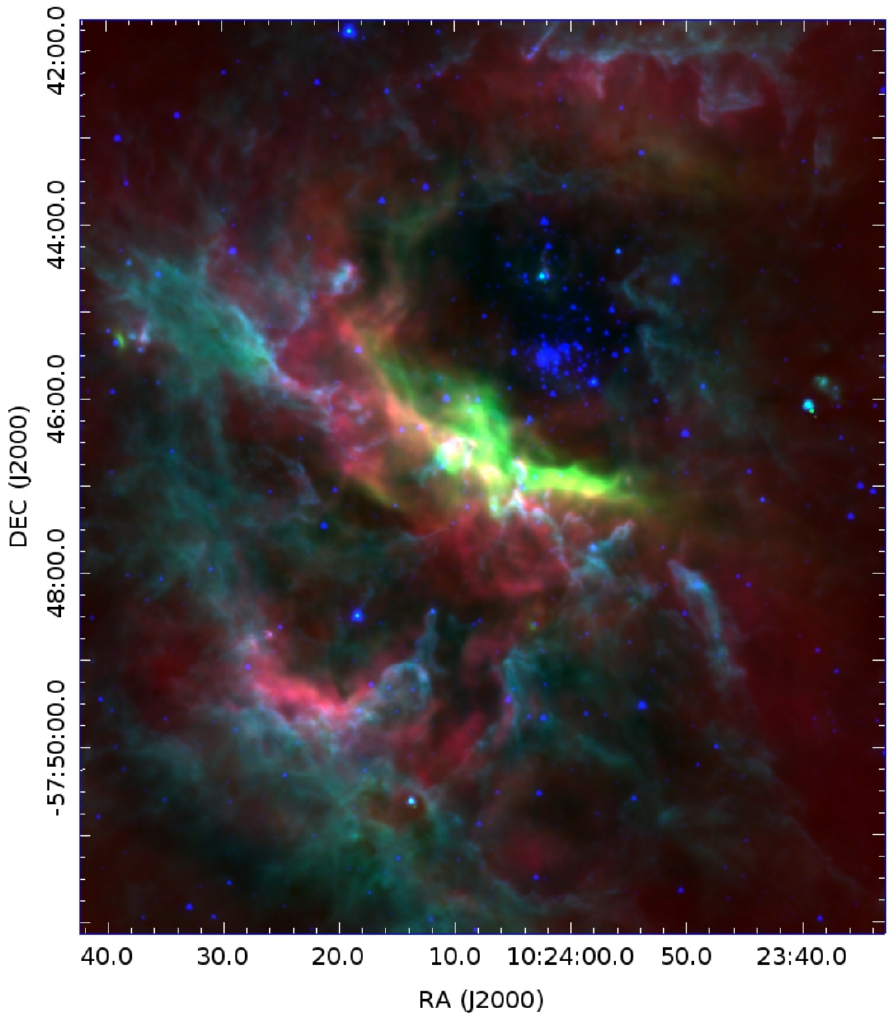}
  \caption{RCW~49 field as seen in the radio continuum (9 GHz, in red) and Spitzer-GLIMPSE band 1 (3.6~$\mu$m) in blue and band 4 (8~$\mu$) in green \citep{paula2013}.}
  \label{figure8}
\end{figure}

Since radio interferometric observations provide a very high angular resolution, the products they deliver can be complemented by others at shorter wavelengths. For example, processes that take place in the interstellar medium are studied by combining radio and infrared data. This is the case of the bright HII region RCW~49 and its very rich ionizing cluster Westerlund~2. Figure~\ref{figure8} shows the distribution of gas, dust, and stars along the field, at arcsec angular scales. The gas is probed by data from the Australia Telescope Compact Array, represented by a 40-pointing mosaic observation; the dust and stars by Spitzer images \citep{paula2013}. There are HE sources in the field, imaged by means of H.E.S.S. and Fermi LAT data. The work discusses possible radio counterparts.

Nearly 200~h of observing time were devoted to the survey of the Cygnus OB2 association and its surroundings with the Giant Metrewave Radio Telescope \citep[see][and related papers]{paula2020}. The observations were made at two MHz bands. About 4000 sources were detected and catalogued (1000 in the two bands), some of them for the first time. The database allowed further studies of the massive early-type stars, protoplanetary disks, young stellar objects, double-lobed objects, and counterparts to unidentified high-energy sources. 

Observations with instruments that record emission outside the radio window are also carried out for projects leaded by IAR researchers, for instance, using XMM-Newton, Chandra, NuSTAR. An example is described in the work by \citet{enzo2022} on the binary  source OAO~1657--415 -- an accreting X-ray pulsar with a high-mass companion --. The authors identified pulsations in NuSTAR data and explained their origin and characteristics, estimated the value of the dipolar magnetic field at the pulsar surface and a obtained a lower limit on the distance of the source.

\section{Scientific research with MIA}

The Multipurpose Interferometric Array, in its full configuration, is expected to be formed at least by 32 elements/antennas with 5-m diameter dishes; an expansion to 64 antennas is also planned \citep[see full description in][this volume]{gancio2023}. The largest 55~km baseline will provide an angular resolution close to 1~arcsec in L-band. The  final coverage is expected to be between 100~MHz and 2.3~GHz. 

With the above parameters, MIA observations can contribute to advancing studies on four major topics: transient sources and timing measurements, sources of non-thermal radiation, neutral hydrogen, from rest to redshifted velocities, and astrophysical plasmas. 

The high-precision timing settings will allow the detection of transient counterparts of gamma-ray bursts, the study of fast radio bursts, pulsars and gravitational waves, and the observation of flares from magnetars. 

The short frequencies at which the MIA receivers will operate are ideal to probe sources where non-thermal radiation is important. This, combined with MIA's high temporal resolution capabilities, will make the instrument optimal for studying the counterparts of unidentified gamma-ray sources, performing multifrequency studies of AGN variability, spectro-temporal studies of X-ray and gamma-ray binaries, studies of the morphology and spectral distribution of supernova remnants, the mapping of continuum non-thermal extended sources, and a long list of other objects.

MIA's high bandwidth, from GHz to a few MHz is well suited to observe the HI line of atoms at rest, but also at large distances, i.e. high redshift sources, with a resolution down to 1 arcsec. This will allow studies of nearby and/or extended galaxies, of the interstellar medium with high angular resolution, and, under certain conditions, of HI at cosmological distances to characterize the Epoch of Reionization.

Emission from astrophysical plasmas is relatively strong even at short centimeter wavelengths. MIA will be very useful for the physical and kinematical characterization of HII regions, for the study of the OH maser variability in star-forming regions and evolved massive stars, and star forming regions in general.

\end{document}